\documentclass[10pt,twoside,twocolumn]{book}

\usepackage[T1]{fontenc}
\usepackage[latin1]{inputenc}
\usepackage{geometry}
\usepackage{amssymb,amsmath,amsfonts}
\usepackage{nicefrac}
\usepackage{latexsym}
\usepackage{fancyhdr}
\usepackage{graphicx}
\usepackage{amsthm,multibox}
\usepackage{color}
\usepackage{hhline,xspace,rotate}
\usepackage{epsfig}
\usepackage{psfig}
\usepackage{pst-all}
\usepackage{pst-poly}
\usepackage{pst-coil}
\usepackage{multido}
\usepackage{url}
\usepackage[french]{varioref}
\usepackage{colortbl}
\usepackage{verbatim}

\usepackage{import}
\usepackage{setspace}
\usepackage[breaklinks=true,debug=true,ps2pdf]{hyperref}

\makeatletter

\newcommand \thearabicpart {\@arabic\c@part} 

\newlength{\picheight}
\newlength{\titlewidth}
\newsavebox{\picbox}
\savebox{\picbox}{%
  \parbox[b][3.6cm][c]{2.4cm}{}%
}
\newcommand{\makepicbox}[1]{%
  \savebox{\picbox}{\includegraphics[height=\picheight]{#1}}}
\def\institute#1{\gdef\@institute{#1}}
\def\@institute{\@latex@warning@no@line{No \noexpand\institute given}}
\def\maketitle{%
\refstepcounter{chapter}%
\twocolumn[
\null
\setlength{\titlewidth}{\textwidth}
\addtolength{\titlewidth}{-2\wd\picbox}
\addtolength{\titlewidth}{-3em}
\hskip\wd\picbox%
\hfill{\large\bf
  \parbox[b]
{\titlewidth}{%
    \centering\@title\\[2ex plus 2 fill]
    \Large\@author\\[0ex plus 1 fill]
    \large\it\@institute%
  }%
}%
\hfill\usebox\picbox
\par%
\vskip 2em%
]%
\thispagestyle{plain}%
\addcontentsline{toc}{chapter}{{\let\\\protect\xspace\@author: \itshape\@title}}%
\setcounter{section}{0}%
}

\def\@part[#1]#2{%
    \ifnum \c@secnumdepth >-2\relax
      \refstepcounter{part}%
      \addcontentsline{toc}{part}{\thepart\hspace{1em}#1}%
    \else
      \addcontentsline{toc}{part}{#1}%
    \fi
    \def\rjcpartname{#1}
    \markboth{}{}%
    {\centering
     \interlinepenalty \@M
     \normalfont
     \ifnum \c@secnumdepth >-2\relax
       \huge\bfseries\sffamily \partname\nobreakspace\thepart
       \par
       \vskip 20\p@
     \fi
     \Huge \bfseries #2\par}%
    \@endpart}

\def\@lbibitem[#1]#2{%
  \@skiphyperreftrue
  \H@item[%
    \ifx\Hy@raisedlink\@empty
      \hyper@anchorstart{cite.\thechapter.#2}\@BIBLABEL{#1}\hyper@anchorend
    \else
      \Hy@raisedlink{\hyper@anchorstart{cite.\thechapter.#2}\hyper@anchorend}%
      \@BIBLABEL{#1}%
    \fi
    \hfill
  ]%
  \@skiphyperreffalse
  \if@filesw
    \begingroup
      \let\protect\noexpand
      \immediate\write\@auxout{%
        \string\bibcite{\thechapter.#2}{#1}%
      }%
    \endgroup
  \fi
  \ignorespaces
}%
\def\@bibitem#1{%
  \@skiphyperreftrue\H@item\@skiphyperreffalse
  \Hy@raisedlink{\hyper@anchorstart{cite.\thechapter.#1}\relax\hyper@anchorend}%
  \if@filesw
    \begingroup
      \let\protect\noexpand
      \immediate\write\@auxout{%
        \string\bibcite{\thechapter.#1}{\the\value{\@listctr}}%
      }%
    \endgroup
  \fi
  \ignorespaces
}%
\def\@citex[#1]#2{%
  \let\@citea\@empty
  \@cite{\@for\@citeb:=#2\do
    {\@citea\def\@citea{,\penalty\@m\ }%
     \edef\@citeb{\thechapter.\expandafter\@firstofone\@citeb\@empty}%
     \if@filesw\immediate\write\@auxout{\string\citation{\@citeb}}\fi
     \@ifundefined{b@\@citeb}{\mbox{\reset@font\bfseries ?}%
       \G@refundefinedtrue
       \@latex@warning
         {Citation `\@citeb' on page \thepage \space undefined}}%
       {\hbox{\csname b@\@citeb\endcsname}}}}{#1}}

\renewenvironment{thebibliography}[1]
     {\section*{\refname}%
      \@mkboth{\MakeUppercase\refname}{\MakeUppercase\refname}%
      \list{\@biblabel{\@arabic\c@enumiv}}%
           {\settowidth\labelwidth{\@biblabel{#1}}%
            \leftmargin\labelwidth
            \advance\leftmargin\labelsep
            \@openbib@code
            \usecounter{enumiv}%
            \let\p@enumiv\@empty
            \renewcommand\theenumiv{\@arabic\c@enumiv}}%
      \sloppy
      \clubpenalty4000
      \@clubpenalty \clubpenalty
      \widowpenalty4000%
      \sfcode`\.\@m}
     {\def\@noitemerr
       {\@latex@warning{Empty `thebibliography' environment}}%
      \endlist}

\makeatother

\usepackage[frenchb, british]{babel}
\newcommand*{\refname}{Bibliography}

\begin{document}

\selectlanguage{frenchb}


%

%
  {
    \cleardoublepage
    \makepicbox{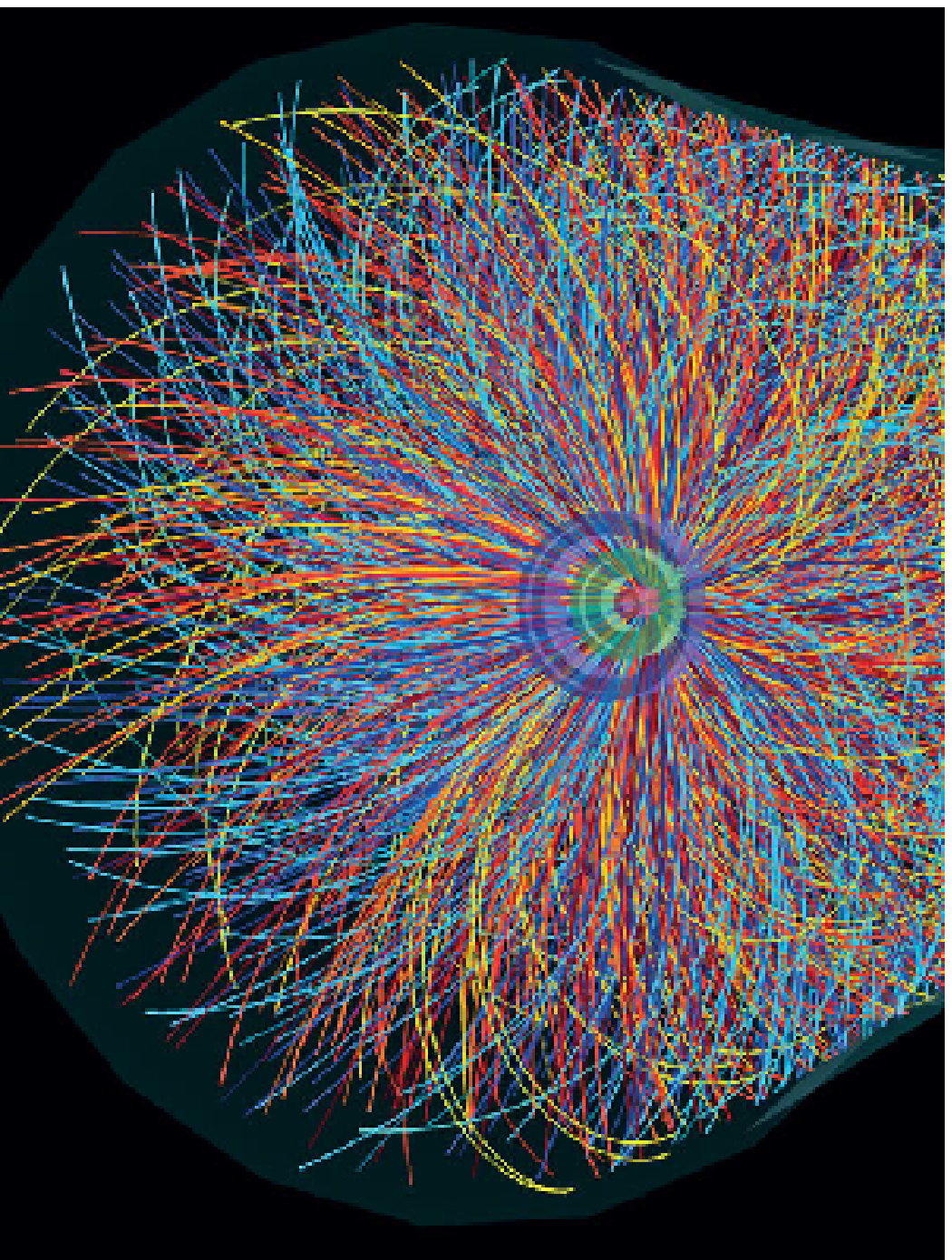}
    \begin{otherlanguage}{british}


\institute{IPHC Strasbourg, groupe ALICE}
\title{Introduction to Quark-Gluon Plasma session}

\author{Antonin MAIRE}

\maketitle


In the following, a brief introduction to the physics of Quark-Gluon Plasma (QGP) will be presented, 
in order to set the stage for the corresponding session proceedings\footnotemark.

\footnotetext{The choice has been made not to provide the corresponding overview on hadronic physics,
being yet the first half of the current session.
(Hadronic physics deals with the various tests to probe the inner structure of nucleons, meant to derive 
distributions of electric charges, spins, ... over the partons defining the sea.)
No misunderstanding here~:
this is not motivated by any bitterness towards this physics (!)
but simply by the blatant lack of knowledge of the session convenor in this area.}

\section{Bjorken scenario of a heavy-ion collision}
\label{sec:Maire-BjorkenScenario}

Under sufficiently high energy densities, it becomes possible to create an exotic state of partonic matter, 
the so-called Quark-Gluon Plasma. 
The QGP consists in a high-density medium of strongly interacting matter, 
made of quarks and gluons which are deconfined (in contrast with hadrons of ordinary matter) 
and thermalised (temperature above the critical temperature of 155-160~MeV~$\simeq 10^{12}$~K, 
as predicted by Lattice QCD \cite{philipsen2013reviewEOSinLQCD}).

Experimentally, this can be enacted by colliding nuclear matter, especially via collisions of heavy ions (\mbox{A-A}), that are
typically $^{63}_{29}$Cu, $^{197}_{~79}$Au, $^{238}_{~92}$U at RHIC \cite{rhic2014runOverview}
and $^{208}_{~82}$Pb at the LHC \cite{lhc2014coordinationPage}. 
With the ultra-relativistic energies achieved at those two colliders 
(centre-of-mass energy, $\sqrt{s_\text{\textsc{nn}}} \leq$~0.2~TeV at RHIC; 
$\sqrt{s_\text{\textsc{nn}}} \leq$~5.52~TeV at the LHC), 
the intent is to form in the laboratory a plasma that should be similar to the state of the primordial universe, 
in place up to a few microseconds after the Big Bang.

Such a \emph{laboratory} QGP occurs in the case there is enough energy deposited into the fireball, 
\emph{i.e.} when enough nucleons and their innner constituants (the aforementioned partons) participate to the initial interaction. 
To quantify how much head-on the collision is, one traditionnally refers to the MC Glauber model \cite{miller2007GlauberModelingInHIC},
model that can be related to some experimental observables (see e.g. \cite{alice2013centralityDeterminationInPbPb2pt76TeV}).
In short, the model provides a way to classify collisions, in terms of \emph{centrality} : 
the more central the collision, the smaller the impact parameter between the centres of the two incident nuclei 
and finally, the higher the released energy.
In the event of a sufficiently \emph{central} collision, a QGP can be produced,
following the schematic scenario originally outlined by J.D. Bjorken \cite{bjorken1983hrn}. 
Such a scenario is depicted by \textsc{f}igs.~\ref{Fig:Maire-Fig1-BjorkenScenario-Madai} and \ref{Fig:Maire-Fig2-BjorkenScenario}. 
The \textsc{f}ig.~\ref{Fig:Maire-Fig1-BjorkenScenario-Madai} gives a visualisation of the collision evolution itself 
(initial nucleons, then deconfined partons, before final-state hadrons) 
while \textsc{f}ig.~\ref{Fig:Maire-Fig2-BjorkenScenario} is a sketch of the "temperature" evolution.

    \begin{figure*}[!htb]
                \centering
                \includegraphics[width=1.00\textwidth, angle=0, clip=true, trim=0cm 0 0 0]{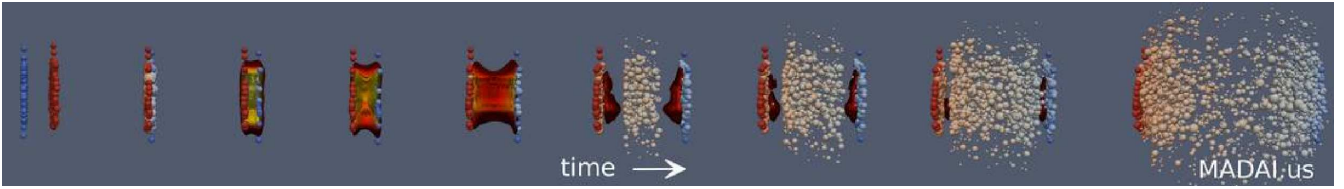}
        \caption{Snapshot of a Bjorken scenario simulated for a heavy-ion collision, leading to the production of a QGP.
                 Courtesy of the MADAI project \cite{madai2014heavyIon}.}
        \label{Fig:Maire-Fig1-BjorkenScenario-Madai}
    \end{figure*}

    \begin{figure*}[!htb]
                \centering
                \includegraphics[width=0.9\textwidth, angle=0, clip=true, trim=0cm 0 0 10cm]{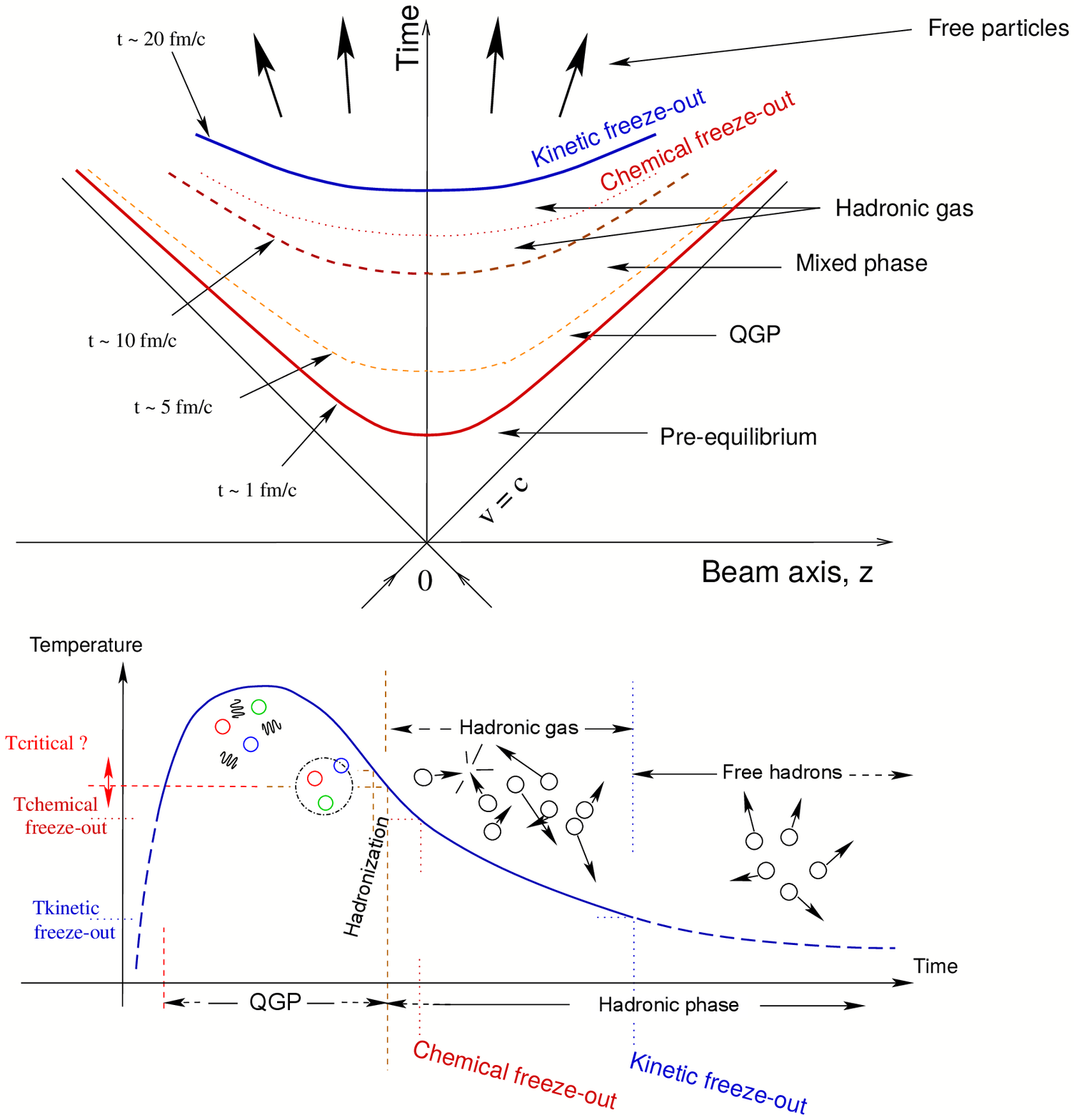}
        \caption{Sketch of the Bjorken scenario for a heavy-ion collision (Fig.~I.10 in \cite{maire2011phdThesis}).}
        \label{Fig:Maire-Fig2-BjorkenScenario}
    \end{figure*}

When the nuclei meet, the formed system starts with a pre-equilibrium stage
in which deposited energy starts to create (anti)quarks and gluons that will be the bulk of the parton population.
Due to the initial longitudinal boost inherited from the incident nuclei, superimposed to the radial explosive behaviour, 
the system is in cylindrical expansion; 
its volume will keep increasing up to the very last moments of the interactions.
After a short span\footnotemark{} of $\precsim 0.5-1$ fm/$c$, 
the medium reaches a thermodynamic equilibrium at which a temperature can be defined.
At that time, the volume is still small enough to have a very high energy density
($\pi r_{Pb}^2 c \tau_0 \simeq 10^2 ~\mathrm{fm}^3$ 
(eq.2 of Bjorken formula in \cite{cms2012transvEnergyInPbPb2pt76TeV}), 
with the radius of the colliding Pb nuclei being $r_{Pb}~\approx$~7~fm and the equilibration time, $\tau_0
\simeq$ 1 fm/$c$) 
and undergo the phase transition towards the plasma.
Indeed, the critical density from Lattice QCD can be roughly estimated to be $\succsim$~1~GeV/fm$^3$, 
\emph{i.e.} above the energy density of a confined nucleon of $\approx$ 1~fm$^3$ and of mass energy of 1~GeV;
the density assessed with heavy-ion data, for $\tau_0~\simeq$~1~fm/$c$, 
is then of the order of 14~GeV/fm$^3$  at the LHC \cite{cms2012transvEnergyInPbPb2pt76TeV}).

\footnotetext{In the following, typical figures will be given for LHC energies; numbers may differ a bit for RHIC ones.}
        
With a constant energy but an increasing volume, the medium will inescapably cool down.
The lowering of the temperature is marked by two key instants :
\begin{enumerate}
  \item after some 7-10 fm/$c$ \cite{song2015vishnuPredictionsForpiKpLambdaXiOmegaAtLHC}, 
        the vast majority of the fireball has turned back into a hadronic gas, 
        the overall system has begun to hadronise and to go back to confined matter. 
        However the formed hadrons can still interact inelastically. The chemical composition changes continuously.
        Such interactions go on, down to the critical temperature of the \emph{chemical freeze-out},
        \emph{i.e.} down to the moment the hadronic species and their respective populations become fixed 
        (e.g. the abundances of $\mathrm {\pi^-}$, p, $\mathrm{\Xi}$... become final). 
        At that stage, the system now typically covers $\mathcal{O}(10^3$ fm$^3)$ \cite{floris2014StatHadroModelQM2014}.
  \item Inelastic collisions cease but there still exist elastic collisions between hadrons. 
        Momentum distributions of given species can evolve.
        Such momentum transfers are possible down the \emph{kinetic freeze-out}, 
        occurring at a typical time of $\succsim$ 10 fm/$c$ \cite{alice2011hbtPionsInPbPb2pt76TeV} $\approx 10^{-23}$ s. 
        After this ultimate freeze-out, the system vanishes into free hadrons that fly into the detection apparatus.
\end{enumerate}

The last time estimate that is mentioned should be noted.
The consequence of this is that experimentalists are irrevocably condemned to never observe a QGP live :
any observation, even with the fastest detectors and electronic readout ($\succsim 10^{-12}$ s), 
will take place at a time when the QGP will have already disappeared.
The experimental study of a QGP with heavy-ion collisions can only consist in observing more or less sheer products,
more or less direct remnants of the plasma stage in the collision evolution.

One can give several examples to illustrate the possible blur of experimental observables.
    Consider final hadrons, they will be formed out of quarks and gluons stemming from the plasma phase,
such baryons and mesons should reflect the experience of those former partons;
however the re-interactions in the hadronic phase, in between QGP and detection stages, 
can certainly intervene and partially distort the final picture.
    Another example can be quoted with electromagnetic probes like photons : they should leave the medium unaffected 
(being insensitive to the strongly interacting medium) 
but the exact time at which a given photon of a given momentum has been produced is not accurately accessible 
while the conditions in the medium (its temperature, its parton density, its interaction density, ...) are changing rapidly...
--~These are only two examples among many others but are just there with the intent to suggest the reader that
further subtilities should be borne in mind while discussing QGP physics.

\section{What is a \emph{hard probe} ?}
\label{sec:Maire-WhatIsAHardProbe}

The scenario described in the previous section is likely too simplistic when compared to what may happen effectively in collisions 
For instance, one can imagine that any hadron of any type is 
either stabilised exactly at the same time of a precise and sudden chemical freeze-out
or that this chemical freeze-out is not a unique moment but has some time span, 
each species population being final at contiguous thus different moments...
- This being said, as simplist as the scenario may look like, it nonetheless allows us to get the big picture and, 
for what is next, better settle which notions a \emph{hard probe} recovers in the context of QGP physics.

A hard probe is defined as a particle stemming from energetic processes (hard scatterings),
in essence only possible at the early collision stage 
(\emph{i.e.} from the initial interactions between incoming quarks or gluons), 
and later, "crossing" the medium and experiencing the whole story of it.
Hard probes are often also referred to as "penetrating probes" or "tomographic probes" of the medium.

In concrete terms, the list of hard probes boils down to :
    \begin{enumerate}
    \itemsep-0.2em 
        \item charm quarks
        \item beauty  quarks
        \item high-$p_{\text{\textsc{t}}}$ particles and jets
        \item electro-weak bosons (W, Z, photons)
    \end{enumerate}

In the following proceedings, for what concerns the ALICE contributions of the session, it will deal with some of these items, 
focusing essentially on high-energy photons and charm hadrons.

\end{otherlanguage}
  }

\end{document}